\documentclass[preprint,secnumarabic,amssymb, nobibnotes, aps, prf, superscriptaddress]{revtex4-2}

\usepackage{amsmath}
\usepackage{graphicx}

\usepackage{url}
\usepackage{soul}
\usepackage{xcolor}
\usepackage{longtable}
\usepackage{multirow,booktabs}
\usepackage{threeparttable}

\begin{document}

\title{Pore-scale statistics of temperature and thermal energy dissipation rate in turbulent porous convection}

\author{Ao Xu}%
\affiliation{School of Aeronautics, Northwestern Polytechnical University, Xi'an 710072, China}
\affiliation{Institute of Extreme Mechanics, Northwestern Polytechnical University, Xi'an 710072, China}

\author{Ben-Rui Xu}%
\affiliation{School of Aeronautics, Northwestern Polytechnical University, Xi'an 710072, China}

\author{Heng-Dong Xi}
\email[Author to whom correspondence should be addressed: ]{hengdongxi@nwpu.edu.cn}
\affiliation{School of Aeronautics, Northwestern Polytechnical University, Xi'an 710072, China}
\affiliation{Institute of Extreme Mechanics, Northwestern Polytechnical University, Xi'an 710072, China}

\date{\today}%

\begin{abstract}
We report pore-scale statistical properties of temperature and thermal energy dissipation rate in a two-dimensional porous Rayleigh-B\'enard (RB) cell.
High-resolution direct numerical simulations were carried out for the fixed Rayleigh number ($Ra$) of $10^{9}$ and the Prandtl numbers ($Pr$) of 5.3 and 0.7.
We consider sparse porous media where the solid porous matrix is impermeable to both fluid and heat flux.
The porosity ($\phi$) range  $0.86 \leq \phi \le 0.98$, the corresponding Darcy number ($Da$) range $10^{-4}<Da<10^{-2}$ and the porous Rayleigh number ($Ra^{*}=Ra\cdot Da$) range $10^{5} < Ra^{*} < 10^{7}$.
Our results indicate that the plume dynamics in porous RB convection are less coherent when the solid porous matrix is impermeable to heat flux, as compared to the case where it is permeable.
The averaged vertical temperature profiles remain almost a constant value in the bulk, while the mean-square fluctuations of temperature increases with decreasing porosity.
Furthermore, the absolute values of skewness and flatness of the temperature are much smaller in the porous RB cell than in the canonical RB cell.
We found that intense thermal energy dissipation occurs near the top and bottom walls, as well as in the bulk region of the porous RB cell. In comparison with the canonical RB cell, the small-scale thermal energy dissipation field is more intermittent in the porous cell, although both cells exhibit a non-log-normal distribution of thermal energy dissipation rate.
This work highlights the impact of impermeable solid porous matrices on the statistical properties of temperature and thermal energy dissipation rate,
and the findings may have practical applications in geophysics, energy and environmental engineering, as well as other fields that involve the transport of heat through porous media.

\footnote{
This article may be downloaded for personal use only.
Any other use requires prior permission of the author and APS.
This article appeared in Xu \emph{\emph{et al.}}, Phys. Rev. Fluids \textbf{8}, 093504 (2023) and may be found at \url{https://doi.org/10.1103/PhysRevFluids.8.093504}.
}
\end{abstract}

\maketitle

\section{\label{Section1}Introduction}

Thermal convection in porous media is frequently encountered in geophysics, energy and environmental engineering, and so on \cite{huppert2014fluid,de2016influence,amooie2018solutal}.
An example is geothermal energy, which involves the extraction of thermal energy from the earth's crust \cite{barbier2002geothermal}.
Specifically, the heat generated and stored in the earth warms water that has infiltrated underground reservoirs, and the hot water can escape to the surface as steam.
Another example is redox flow battery \cite{xu2017lattice}, which is an energy storage device used to store intermittent solar and wind power.
The performance of the redox flow batteries relies on the coupled transport of electrolyte, heat, mass, and electrons in the porous electrodes.
In thermal convection \cite{lohse2010small,chilla2012new,xia2013current,xia2023tuning,verma2018physics,wang2020vibration}, the key parameter quantifying the strength of buoyancy forces over dissipation force is the Rayleigh number $Ra=\beta g\Delta_{T} H^{3} /(\nu \alpha)$.
Here, $\beta$, $\alpha$, and $\nu$ are the thermal expansion coefficient, thermal diffusivity, and kinematic viscosity of the fluid, respectively;
$g$ is the gravitational acceleration, and $\Delta_{T}=T_{hot}-T_{cold}$  is the temperature difference across the fluid layer of height $H$.
For the porous media, its ability to transmit fluids is quantified by the permeability $K$, and it is usually represented by the dimensionless Darcy number as $Da=K / L^{2}$, where $L$ is the characteristic length.
The permeability of the porous media is only determined by the geometry of the porous structure, and its value is a complex function of various parameters including the porosity $\phi$ (i.e., the fluid volume fraction) of the porous media.

Fluid flows and associated transport processes in porous media are complex phenomena that can occur over a wide range of spatial and temporal scales.
To simulate these phenomena, numerical methods can be classified into two categories, namely, the representative elementary volume (REV)-scale method and the pore-scale method \cite{whitaker1998method,davit2013homogenization}.
The REV-scale method considers volume-averaged flow quantities (such as velocity, pressure, and permeability) over a representative volume that consists of many pores.
Empirical relations, such as the Blake-Kozeny-Carman relation  \cite{bird2006transport}, can be used to efficiently estimate permeability $K$ of the porous structure.
For porous media convection, the Darcy-Oberbeck-Boussinesq (DOB) equations can be derived using the volume-averaged approach \cite{nield2006convection}.
While the REV-scale method has the advantage of high computational efficiency, its accuracy relies heavily on the adopted empirical relations.
A review paper by Hewitt \cite{hewitt2020vigorous} provides an in-depth exploration of the REV-scale modeling and simulation of convection in porous media.
In contrast, the pore-scale method resolves the geometry of individual pores, allowing for the calculation of constitutive closure relations such as permeability as a function of porosity.
This method can accurately reflect the geometrical effect of porous structure on the transport process; however, the high computational cost of pore-scale simulation limits its wide engineering applications.
In short, both the REV-scale method and the pore-scale method have their respective advantages and limitations.
Choosing the appropriate method depends on the specific requirements of the problem at hand, including the desired level of accuracy and computational resources available \cite{mattila2016prospect,succi2020toward}.

In turbulent thermal convection, to quantify the dissipation of thermal energies due to thermal diffusivity, the thermal energy dissipation rate is defined as $\varepsilon_{T}(\mathbf{x}, t)=k \sum_{i}\left[\partial_{i} T(\mathbf{x}, t)\right]^{2}$.
In the canonical Rayleigh-B\'enard (RB) convection cell filling with pure fluid (i.e., a fluid layer heated from the bottom and cooled from the top), Shraiman and Siggia \cite{shraiman1990heat} derived exact relations of global average $\varepsilon_{T}=\left\langle\varepsilon_{T}(\mathbf{x}, t)\right\rangle_{V}=\kappa \Delta^{2}_{T} L^{-2} Nu$,
which further form the backbone of the Grossman-Lohse (GL) theory \cite{grossmann2000scaling,grossmann2004fluctuations} on turbulent heat transfer.
With the aid of DNS results, Emran and Schumacher \cite{emran2008fine} analyzed the probability density functions (PDFs) of $\varepsilon_{T}$ in a cylindrical cell.
They found the PDFs deviate from a log-normal distribution but fit well by a stretched exponential distribution, which is similar to passive scalar dissipation rate in homogeneous isotropic turbulence.
Subsequently, Kaczorowski and Wagner \cite{kaczorowski2009analysis} analyzed the contributions of bulk and boundary layers and plumes to the PDFs of the $\varepsilon_{T}$ in a rectangular cell, and they found that the core region scaling changes from pure exponential to a stretched exponential scaling as $Ra$ increases.
Recently, Xu \emph{et al.} \cite{xu2019statistics}, Zhang \emph{et al.} \cite{zhang2017statistics}, and Bhattacharya \emph{et al.} \cite{bhattacharya2019scaling} obtained the $Ra$ scaling relations for the thermal dissipation rate in the bulk and the boundary layers at low-, moderate-, and high-$Pr$ regime, respectively.
An interesting finding is that despite the boundary layer region occupied a much smaller volume, the globally averaged thermal energy dissipation rate from the boundary layer region is still larger than that from the bulk region.

Although considerable efforts have been devoted to exploring the statistical properties of temperature and thermal energy dissipation rate in the canonical RB convection cell, fewer studies have focused on investigating the pore-scale statistics of these quantities in the turbulent porous RB convection cells \cite{liu2020rayleigh,gasow2020effects,liu2021lagrangian,korba2022effects}.
In a study by Liu \emph{et al.} \cite{liu2020rayleigh}, a porous RB cell was considered where the thermal properties of the fluid and solid phases were assumed to be the same, indicating that the solid porous media were permeable to heat flux.
In contrast, in our work, we assume that the solid matrix is impermeable to heat flux, which is a reasonable assumption for porous media with much lower thermal conductivities compared to that of the fluid.
For example, porous structures made of ceramics (alumina and zirconia) and carbon-based materials (activated carbon and carbon nanotubes) serve as effective thermal insulators.
Moreover, because of the analogy between heat transfer and mass transfer, the impermeable heat flux boundary condition can be considered analogous to non-reacting boundary conditions at the surface of solid obstacles \cite{schlander2022analysis};
as a result, our work on scalar transport could potentially inspire further research on convective mass transfer in porous media \cite{buaria2022intermittency}.
The rest of this paper is organized as follows.
In Sec. \ref{Section2}, we present the numerical details including pore-scale simulation of fluid flows and heat transfer, as well as generation and characterization of porous media.
In Sec. \ref{Section3}, we first present general features of fluid flows and heat transfer in the pores, and then we analyze statistics of temperature and thermal energy dissipation rate.
In Sec. \ref{Section4}, the main findings of the present work are summarized.

\section{\label{Section2}Numerical methods}

\subsection{Mathematical model for fluid flows and heat transfer at pore-scale }

In the pore-scale method, individual pore geometry is directly resolved, thus, the governing equations for fluid flows and heat transfer in the pores are the Navier-Stokes equations with Boussinesq approximation:
\begin{subequations}
\begin{equation}
\nabla \cdot \mathbf{u}_{f}=0
\end{equation}
\begin{equation}
\frac{\partial \mathbf{u}_{f}}{\partial t}+\mathbf{u}_{f} \cdot \nabla \mathbf{u}_{f}=-\frac{1}{\rho_{0}} \nabla P+ \nu_{f} \nabla^{2} \mathbf{u}_{f}+g \beta\left(T_{f}-T_{0}\right) \hat{\mathbf{y}}
\end{equation}
 \begin{equation}
 \frac{\partial T_{f}}{\partial t}+\mathbf{u}_{f} \cdot \nabla T_{f}=\alpha_{f} \nabla^{2} T_{f}
 \end{equation}
\label{Eq.NS}
\end{subequations}
Here, the subscript $f$ denotes the fluid phase. $\mathbf{u}_{f}$, $P$, and $T_{f}$ are the fluid velocity, pressure, and temperature in the pores, respectively.
$\rho_{0}$ and  $T_{0}$ are reference density and temperature, respectively.
$g$ is the gravity value and $ \hat{\mathbf{y}}$  is the unit vector in the vertical direction.
In Eq. (\ref{Eq.NS}), all the transport coefficients (i.e., $\nu_{f}$, $\alpha_{f}$, $\beta$) are assumed to be constants.
Using the nondimensional group
\begin{equation}
    \begin{split}
& \mathbf{x}^{*}=\mathbf{x}/H, \ \ \ t^{*}=t/\sqrt{H/(g\beta \Delta_{T})}, \ \ \ \mathbf{u}^{*}_{f}=\mathbf{u}_{f}/\sqrt{g \beta \Delta_{T}H}, \\
& P^{*}=P/(\rho_{0}g\beta \Delta_{T}H), \ \ \ T^{*}_{f}=(T_{f}-T_{0})/\Delta_{T} \\
    \end{split}
\end{equation}
Equation (\ref{Eq.NS}) can be rewritten in dimensionless form as
\begin{subequations}
\begin{equation}
\nabla \cdot \mathbf{u}^{*}_{f} = 0 \label{Eq.divU-dim}
\end{equation}
\begin{equation}
\frac{\partial \mathbf{u}^{*}_{f}}{\partial t^{*}}+\mathbf{u}^{*}_{f}\cdot \nabla \mathbf{u}^{*}_{f}
=-\nabla P^{*}+\sqrt{\frac{Pr}{Ra}} \nabla^{2}\mathbf{u}^{*}_{f}+T^{*}_{f}\hat{\mathbf{y}} \label{Eq.momentum-dim}
\end{equation}
\begin{equation}
\frac{\partial T^{*}_{f}}{\partial t^{*}}+\mathbf{u}^{*}_{f}\cdot\nabla T^{*}_{f}=\sqrt{\frac{1}{Pr Ra}} \nabla^{2}T^{*} _{f} \label{Eq.temperature-dim}
\end{equation}
\end{subequations}
In the following, for convenience, we will drop the superscript star ($*$) to denote a dimensionless variable.
The dimensionless parameters of the Rayleigh number ($Ra$) and the Prandtl number ($Pr$) are defined as
\begin{equation}
Ra=\frac{g\beta \Delta_{T}H^{3}}{\nu_{f} \alpha_{f}}, \ \ \ Pr=\frac{\nu_{f}}{\alpha_{f}}
\end{equation}

\subsection{Lattice Boltzmann model for incompressible thermal flows}

We adopt the lattice Boltzmann (LB) method \cite{maggiolo2023transition,zhou2022neural,ju2022pore} as the numerical tool for direct numerical simulation of turbulent thermal convection in the pores.
In the LB method, to solve Eqs. (\ref{Eq.NS}a) and (\ref{Eq.NS}b), the evolution equation of density distribution function is written as
\begin{equation}
f_{i}\left(\mathbf{x}+\mathbf{e}_{i} \delta_{t}, t+\delta_{t}\right)-f_{i}(\mathbf{x}, t)=-\left(\mathbf{M}^{-1} \mathbf{S}\right)_{i j}\left[\mathbf{m}_{j}(\mathbf{x}, t)-\mathbf{m}_{j}^{(\mathrm{eq})}(\mathbf{x}, t)\right]+\delta_{t} F_{i}^{'}
\label{Eq.f}
\end{equation}
To solve Eq. (\ref{Eq.NS}c), the evolution equation of temperature distribution function is written as
\begin{equation}
g_{i}\left(\mathbf{x}+\mathbf{e}_{i} \delta_{t}, t+\delta_{t}\right)-g_{i}(\mathbf{x}, t)=-\left(\mathbf{N}^{-1} \mathbf{Q}\right)_{i j}\left[\mathbf{n}_{j}(\mathbf{x}, t)-\mathbf{n}_{j}^{(\mathrm{eq})}(\mathbf{x}, t)\right]
\label{Eq.g}
\end{equation}
Here,  $f_{i}$  and $g_{i}$  are the density and temperature distribution functions, respectively.
$\mathbf{x}$ is the fluid parcel position, $t$  is the time,  $\delta_{t}$ is the time step.
$e_{i}$  is the discrete velocity along the $i$ th direction.
$\mathbf{M}$ is a $9 \times 9$ orthogonal transformation matrix based on the D2Q9 discrete velocity model;
$\mathbf{N}$ is a $5 \times 5$ orthogonal transformation matrix based on the D2Q5 discrete velocity model.
The equilibrium moments $\mathbf{m}^{(\text{eq})}$  in Eq. (\ref{Eq.f}) are
\begin{equation}
\mathbf{m}^{(\text{eq})}=\rho\left[1, \ -2+3\left|\mathbf{u}_{f}\right|^{2}, \ 1-3\left|\mathbf{u}_{f}\right|^{2}, \ u_{f},-u_{f}, \ v_{f}, \ -v_{f}, \ 2 u_{f}^{2}-v_{f}^{2}, \ u_{f} v_{f}\right]^{T}
\end{equation}
The equilibrium moments $\mathbf{n}^{(\text{eq})}$  in Eq. (\ref{Eq.g}) are
\begin{equation}
\mathbf{n}^{(\text{eq})}=\left[T_{f}, \ u_{f} T_{f}, \ v_{f} T_{f}, \ a_{T} T_{f}, \ 0\right]^{T}
\end{equation}
where  $a_{T}$ is a constant determined by the thermal diffusivity as $a_{T}=20 \sqrt{3} \kappa_{f}-4$.
The relaxation matrix $\mathbf{S}$ is $\mathbf{S}=\operatorname{diag}\left(s_{\rho}, s_{e}, s_{s}, s_{j}, s_{q}, s_{j}, s_{q}, s_{\nu}, s_{\nu}\right)$, and the kinematic viscosity of the fluids is calculated as $v_{f}=\left(s_{v}^{-1}-0.5\right) / 3$.
The relaxation matrix $\mathbf{Q}$ is given by $\mathbf{Q}=\operatorname{diag}\left(0, q_{k}, q_{k}, q_{e}, q_{v}\right)$, where $q_{\kappa}=3-\sqrt{3}$  and $q_{e}=q_{\nu}=4\sqrt{3}-6$.

The macroscopic fluid variables of density $\rho_{f}$, velocity $\mathbf{u}_{f}$ and temperature $T_{f}$  are calculated as $\rho_{f}=\sum_{i=0}^{8} f_{i}$, $\mathbf{u}_{f}=\left(\sum_{i=0}^{8} \mathbf{e}_{i} f_{i}+\mathbf{F} / 2\right) / \rho_{f}$ and $T_{f}=\sum_{i=0}^{4} g_{i}$, respectively.
More numerical details on the LB method and validation of the in-house code can be found in our previous work \cite{xu2017accelerated,xu2019lattice,xu2023multi}.

\subsection{Boundary conditions at the fluid-solid interface}

We assume the solid matrix is impermeable to both fluid and heat flux. At the fluid-solid interface, the no-slip velocity boundary conditions can be described as $\mathbf{u}_{f} = 0$; while the adiabatic temperature boundary conditions can be described as  $\partial_{\mathbf{n}} T_{f}=0$.
In the LB method, the-above fluid-solid interface conditions can be mimic by the bounce-back rules for the density and temperature distribution functions as $f_{\bar{i}}\left(\mathbf{x}_{f}, t+\delta_{t}\right)=f_{i}^{*}\left(\mathbf{x}_{f}, t\right)$  and $g_{\bar{i}}\left(\mathbf{x}_{f}, t+\delta_{t}\right)=g_{i}^{*}\left(\mathbf{x}_{f}, t\right)$, respectively.

\subsection{Generation and characterization of porous structure}

\begin{figure}
\centering
\includegraphics[width=0.7\textwidth]{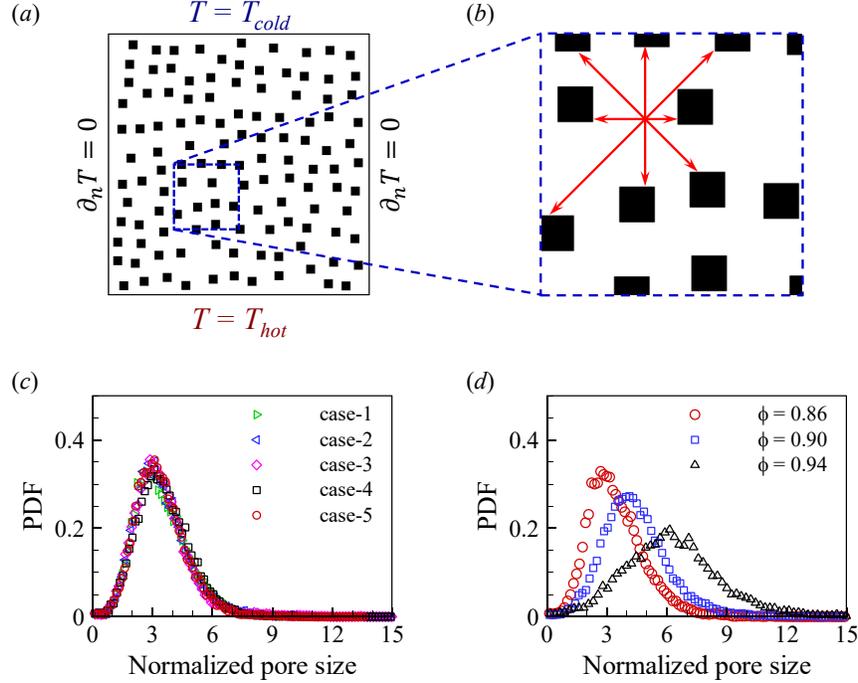}
\caption{ (\textit{a}) Illustration of the porous convection cell. The black region represents the solid matrix, and the white region represents the fluid. (\textit{b}) An enlarged view of the porous region in (\textit{a}), and the schematic illustration of the directional average method to calculate pore size. The probability density functions (PDFs) of normalized pore size for (\textit{c}) different realizations of the porous structure at the same porosity of $\phi=0.86$ and (\textit{d}) porous structure with different porosities.} \label{Fig_demoPorousStructure}
\end{figure}

We artificially construct the porous structure via randomly placing square cylinders of length $d$ in the RB convection cell, as illustrated in Fig. \ref{Fig_demoPorousStructure}(a).
In Fig. \ref{Fig_demoPorousStructure}(b), we further present an enlarged view of the porous region as that in Fig. \ref{Fig_demoPorousStructure}(a), and we illustrate the directional average method \cite{lange2010pore,xu2018lattice} to calculate pore size distribution (PSD).
At each fluid point, we start counting the pore length along with specified directions until reaching a solid point; then, the pore diameter is obtained by averaging the pore length in all given eight directions.
It should be noted that the directional average method provides an approximate assessment of pore size, and it comes with inherent limitations.
This method determines the average span of the pore spaces from a certain point in all possible directions (i.e., eight discrete directions in this work) until an obstacle or the boundary of the domain is encountered.
When assessing fluid points near the boundary, we exclude directions beyond the domain boundary.
The absence of obstacles in certain directions could lead to an unusually large distance, causing an overestimation of pore size.
Nevertheless, the method retains its applicability in this work, as the issues related to boundary effects or statistical anomalies can be mitigated by employing a large computational domain.
In addition, it is particularly beneficial when our goal is to determine the average pore sizes across the porous medium.
Figure \ref{Fig_demoPorousStructure}(c) shows the probability density functions (PDFs) of calculated pore size for five different realizations of the porous structure at the same porosity of  $\phi=0.86$.
Here, the pore size is normalized by the cylinder length $d$.
We can see that despite different realizations of the porous structure generation, the pore size distributions are roughly the same at the same porosity.
In Fig. \ref{Fig_demoPorousStructure}(d), we compare the pore size for porous structure with different porosities of $\phi=0.86$, 0.90 and 0.94.
Overall, the pore size exhibits unimodal distribution, and it increases with the increase of porosity.

\begin{figure}
\centering
\includegraphics[width=0.7\textwidth]{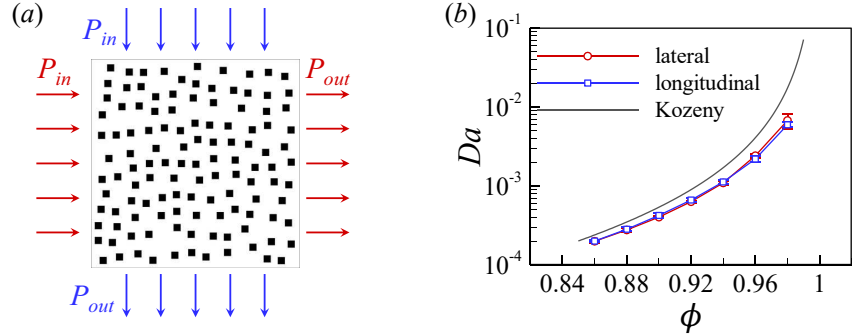}
\caption{ (\textit{a}) Illustration of the simulation settings to calculate the permeability of porous structure; (\textit{b}) Darcy number as a function of porosity (the error bar is calculated based on results from five different realizations of the porous structure at the same porosity).} \label{Fig_permeability}
\end{figure}

To determine the permeability $K$ of the porous structure, we performed another set of pore-scale simulations when the flow is isothermal and in the Darcy regime.
As illustrated in Fig. \ref{Fig_permeability}(a), fluid flows through porous media is driven by a small pressure difference $(P_{in}-P_{out})/P_{ref}=2\times 10^{-4}$ either in lateral or longitudinal direction, such that flow is sufficient slow.
Following the Darcy's law \cite{darcy1856fontaines}, we calculate the permeability tensor as \cite{whitaker1998method}
\begin{equation}
\mathbf{K}=-\mu\frac{\langle \mathbf{u}_{f}  \rangle}{\nabla \langle p \rangle^{f}}
\end{equation}
Here, the intrinsic phase average is defined as $\langle \psi_{f}\rangle^{f}=(1/V_{f})\int_{V_{f}}\psi_{f}dv$,
and the superficial phase average is defined as $\langle \psi_{f} \rangle=(1/V)\int_{V_{f}}\psi_{f}dv$.
$V_{f}$ denotes the volume of the fluid phase within the representative volume $V$, and $\psi_{f}$ is a quantity associated with the fluid phase.
We also check the pore-scale Reynolds number $Re_{D}=\langle u_{f} \rangle^{f} D/\nu$ satisfies $Re_{D} \le O(1)$, thus, the condition to apply Darcy's law is guaranteed.
In Fig. \ref{Fig_permeability}(b), we show the permeabilities in lateral and longitude directions (i.e., $K_{xx}$ and $K_{yy}$), respectively,
which are further presented in terms of dimensionless Darcy number $Da=K/L^{2}$.
Here, the characteristic length $L$ is chosen as convection cell size.
We can see that the lateral and longitudinal permeabilities are generally the same, suggesting the artificially constructed porous structure is homogeneous.
Besides, we compare the calculated permeability with empirical Blake-Kozeny-Carman relation \cite{bird2006transport}
\begin{equation}
K=\frac{\phi^3 D^2}{150(1-\phi)^2}
\end{equation}
We can see from Fig. \ref{Fig_permeability}(b), the empirical relation overestimates the permeability for the sparse porous media, and the deviations increase with increasing the porosity.
For the investigated porosity range $0.86 \le \phi \le 0.98$, the corresponding $Da$ range $10^{-4} < Da < 10^{-2}$.

\subsection{Simulation settings}

We consider a two-dimensional (2D) porous RB convection cell with size $L=H$.
The top and bottom walls of the cell are kept at a constant cold and hot temperature, respectively; the other two vertical walls are adiabatic.
All four walls impose no-slip velocity boundary conditions.
We provide simulation results at Prandtl number of $Pr = 5.3$ and $0.7$ (i.e., corresponding to the working fluid of water and air, respectively) and a fixed $Ra = 10^{9}$.
The porosity ($\phi$) range  $0.86 \leq \phi \le 0.98$, and the corresponding porous Rayleigh number ($Ra^{*}=Ra\cdot Da$) range $10^{5} < Ra^{*} < 10^{7}$, suggesting vigorous convection in porous media \cite{hewitt2020vigorous}.
In addition, we show the scaling of the global quantities on one of the control parameters $Ra$ (for $10^{6} \le Ra \le 10^{9}$), while the porosity is fixed as $\phi = 0.86$ and Prandtl number is fixed as $Pr = 5.3$ and $0.7$.
A total of 100 simulations were carried out for porous convection with impermeable solid matrix, and tabulated values on the results are listed in the Appendix.
For the canonical RB convection, the mesh size of the convection cell is $1024$ l.u.$\times 1024$ l.u.; while for the porous RB convection, the mesh size is even finer with $1200 $ l.u.$\times 1200$ l.u.
The resolution for the cylinder length $d$ is 40 l.u., and the minima gap between two cylinders is 40 l.u.
Here, l.u. denotes the lattice length unit in the LB simulation \cite{huang2009relative}.

For canonical RB convection of pure fluid, we verify the grid spacing $\Delta_{g}$ and time interval $\Delta_{t}$ is properly resolved by comparing with the Kolmogorov and Batchelor scales.
Specifically, the Kolmogorov length scale \cite{kolmogorov1941local} is estimated by the global criterion $\eta_{K}=\left[\nu^{3} /(\varepsilon_{u})_{V, t}\right]^{1 / 4}=H Pr^{1 / 2} /[Ra(Nu-1)]^{1 / 4}$ , the Batchelor length scale \cite{batchelor1959small} is estimated by $\eta_{B}=\eta_{K} Pr^{-1/2}$, and the Kolmogorov time scale \cite{kolmogorov1941local}  is estimated as $\tau_{\eta}=\sqrt{\nu /\left\langle\varepsilon_{u}\right\rangle_{V, t}}=t_{f} \sqrt{Pr /(Nu-1)}$.
Here,  $\varepsilon_{u}$ denotes the kinetic energy dissipation rates, and its global average can be related to the Nusselt number via the exact relation \cite{shraiman1990heat} $\left\langle\varepsilon_{u}\right\rangle_{V, t}=\nu^{3}Ra(Nu-1)/(H^{4}Pr^{2})$.
Simulation results have shown that grid spacing satisfies the criterion of max $\left(\Delta_{g} / \eta_{K}, \ \Delta_{g} / \eta_{B}\right) \leq 0.55$, which ensures spatial resolution; the time intervals are $\Delta_{t} \leq 0.00047 \tau_{\eta}$, thus adequate temporal resolution is guaranteed.
In addition, we validate our results by comparing the Nusselt and Reynolds number with those obtained using the NEK5000 solver (version v19.0) \cite{nek5000},
as well as previous results reported by Zhang \emph{et al.} \cite{zhang2017statistics}.
The tabulated values are presented in Table \ref{tab:label}.
Note that the small deviations in the average statistical value may be attributed to turbulence fluctuations.
For porous RB convection, both the cylinder length ($d = 40$ l.u.) and minima gap between the cylinders (i.e., $40$ l.u.) are much large than that of the boundary layer thickness (around $10$ l.u.), thus the pore space is adequately resolved.

\begin{table}[htbp]
	\centering
    \caption{Heat transfer efficiency and global flow strength in the canonical RB convection.
        The columns from left to right indicate the following:
        the Rayleigh number $Ra$, the Prandtl number $Pr$,
        the Nusselt number $Nu$, and the Reynolds number $Re$ obtained using in-house LB solver, the NEK5000 solver \cite{nek5000}, and previous results reported by Zhang \emph{et al.} \cite{zhang2017statistics}. }
    {
	\begin{tabular}{cccccccc}
		\toprule
		\multirow{2}{*}{$Ra$} & \multirow{2}{*}{$Pr$} & \multicolumn{3}{c}{$Nu$} & \multicolumn{3}{c}{$Re$} \\
		&       & LB     & NEK5000 & Ref. \cite{zhang2017statistics} & LB  & NEK5000 & Ref. \cite{zhang2017statistics}  \\
		\midrule
		$10^6$ \ \   & 5.3  \ \   & 6.93  & 6.92  & 6.87  & 38    & 36    & 38 \\
		$10^6$ \ \   & 0.7  \ \   & 6.33  & 6.31  & 6.30   & 280   & 279   & 279 \\
		$10^7$ \ \   & 5.3  \ \   & 13.36 & 13.25 & 13.28 & 156   & 154   & 156 \\
		$10^7$ \ \   & 0.7  \ \   & 11.42 & 11.36 & 11.37 & 968   & 973   & 968 \\
		$10^8$ \ \   & 5.3  \ \   & 26.23 & 26.25 & 26.21 & 597   & 596   & 596 \\
		$10^8$ \ \   & 0.7  \ \   & 25.32 & 25.25 & 25.25 & 3661  & 3692  & 3662 \\
		$10^9$ \ \   & 5.3  \ \   & 51.50 & 51.34 & 51.28 & 2273  & 2308  & 2269 \\
		$10^9$ \ \   & 0.7  \ \   & 49.75 & 49.76 & 53.51 & 15588 & 15633 & 15101 \\
		\bottomrule
	\end{tabular}}%
	\label{tab:label}%
\end{table}%

\section{\label{Section3}Results and discussion}
\subsection{General fluid flows and heat transfer features in the pores}
A typical snapshot of an instantaneous temperature field in both a porous and a canonical RB convection cell is shown in Fig. \ref{Fig_instant_temperature}, and the corresponding video can be viewed in the Supplemental Material \cite{SM}.
We can see that the flow structure in the porous RB convection exhibits different patterns from that in the canonical RB convection.
In the canonical RB convection [see Figs. \ref{Fig_instant_temperature}(b) and \ref{Fig_instant_temperature}(d)], the rising and falling thermal plumes self-organize into a well-defined large-scale circulation (LSC) that spans the size of the convection cell \cite{xi2004laminar}, and there exist counterrotating corner rolls.
The temperature field is efficiently mixed in the convection cell, with the bulk temperature being almost a constant value of $(T_{hot}+T_{cold})/2$.
In contrast, in the porous RB convection [see Figs. \ref{Fig_instant_temperature}(a) and \ref{Fig_instant_temperature}(c)], the flow structure is less coherent and the large-scale flow circulation is suppressed.
The rising and falling plumes penetrate through the pore throat, resulting in less mixing of the temperature field.
This flow pattern is similar to that observed in the study by Liu \emph{et al.} \cite{liu2020rayleigh}, where the porous matrix was permeable to heat flux.
However, in the current work, we assume that the solid porous matrix is impermeable to heat flux.
When the solid porous matrix does not conduct heat, there is no thermal exchange between the solid and fluid phases, and the fluid phase's ability to effectively interact with the solid phase is diminished compared to scenarios involving a thermally conductive solid porous matrix.
Consequently, the plume dynamics are less coherent than in the previous study.
In the Supplemental Material \cite{SM}, we provide corresponding videos, which allow for a more detailed examination of the flow patterns and temperature field in the two types of convection cells.

\begin{figure}
\centering
\includegraphics[width=0.7\textwidth]{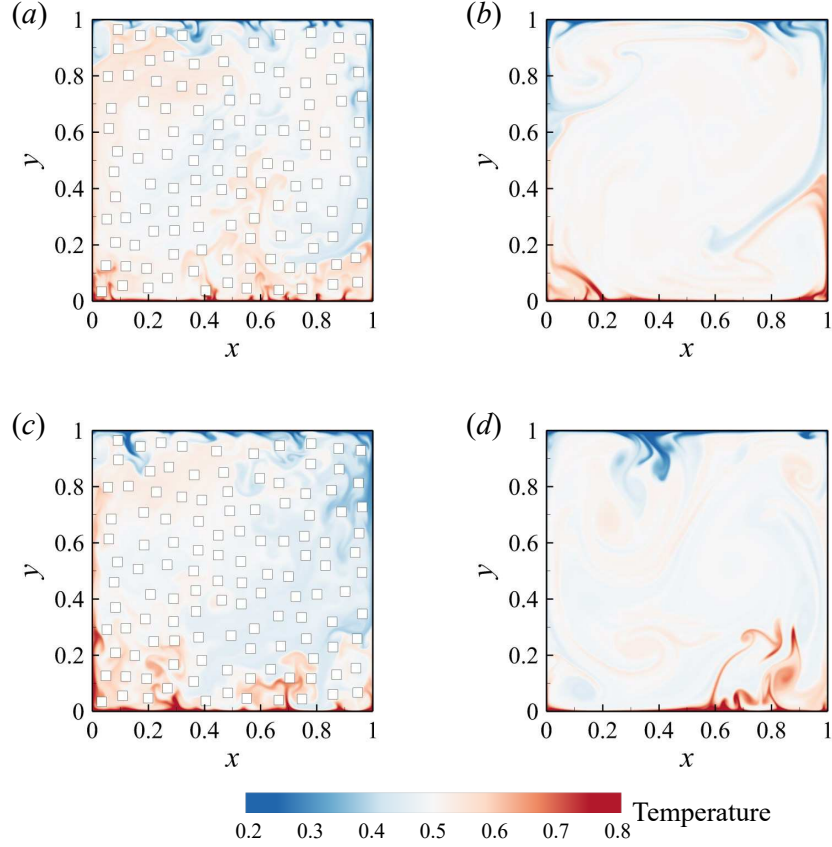}
\caption{A typical snapshot of the instantaneous temperature field for (\textit{a}, \textit{b}) $Pr = 5.3$; (\textit{c}, \textit{d}) $Pr = 0.7$ in (\textit{a}, \textit{c}) a porous RB convection cell with $\phi=0.86$  and (\textit{b}, \textit{d}) a canonical RB convection cell (i.e., $\phi=1.00$).}
\label{Fig_instant_temperature}
\end{figure}

\begin{figure}
\centering
\includegraphics[width=0.88\textwidth]{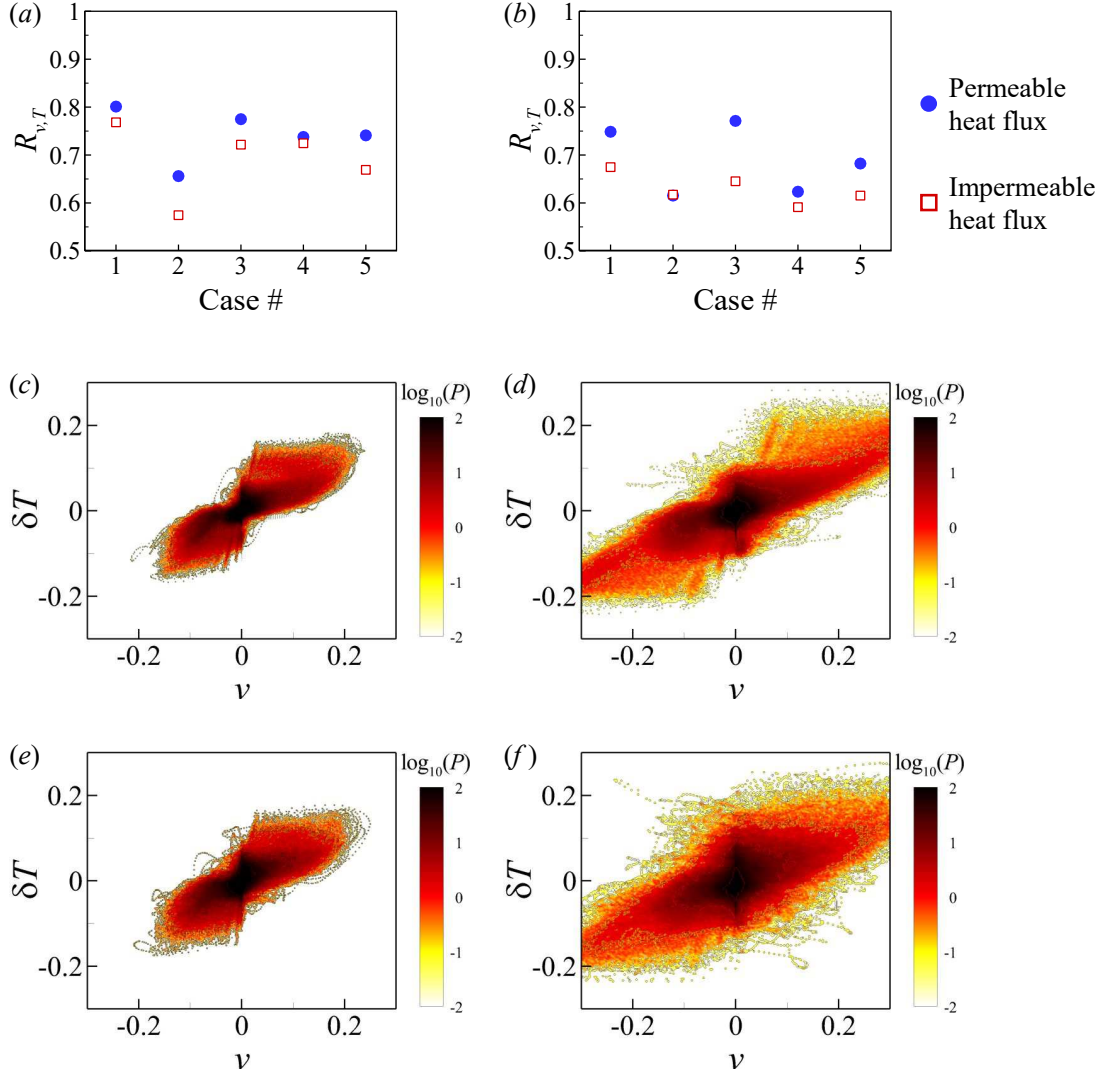}
\caption{(\textit{a}, \textit{b}) Cross-correlation coefficient $R_{v,T}$ between vertical velocity $v$ and temperature $T$ for five different realizations of the porous structure;
(\textit{c}-\textit{f}) logarithmic of joint probability density function (PDF) of vertical velocity $v$ and temperature fluctuation $\delta T$ at the height of $y = 0.5$,
 for (\textit{c},\textit{d}) solid porous matrix being permeable to heat flux, (\textit{e},\textit{f}) solid porous matrix being impermeable to heat flux,
 at $\phi=0.86$, $Ra=10^{9}$, and (\textit{a},\textit{c},\textit{e}) $Pr=5.3$, (\textit{b},\textit{d},\textit{f}) $Pr=0.7$.}
\label{Fig_xcov}
\end{figure}

To validate the above conjecture, we calculate the cross-correlation coefficient between vertical velocity $v$ and temperature $T$ along the mid-plane of the cell, given by
\begin{equation}
R_{v,T}=\frac{\langle [v(t)-\langle v \rangle][T(t)-\langle T \rangle] \rangle}{\sigma_{v}\sigma_{T}}
\end{equation}
where $\sigma_{v,T}$ denotes the standard deviation of $v$ and $T$.
We conducted simulations for two cases:
one with solid porous matrix being permeable to heat flux (referred to as permeable heat flux, following the simulation settings reported by Liu \emph{et al.} \cite{liu2020rayleigh}),
and the other is the solid porous matrix being impermeable to heat flux (referred to as impermeable heat flux).
Five different realizations of the porous structure were considered at the same porosity $\phi=0.86$.
Figures \ref{Fig_xcov}(a) and \ref{Fig_xcov}(b) shows the cross-correlation coefficient $R_{v,T}$ for both cases.
We can observe that $R_{v,T}$ is generally lower for the impermeable heat flux case, which implies that the thermal plumes are less coherent.
We further calculate the joint probability density distribution of the vertical velocity $v$ and temperature fluctuation $\delta T=T-(T_{hot}+T_{cold})/2$.
In comparison to a thermally conductive solid porous medium, an impermeable medium reduces the correlation between vertical velocity and temperature.
This effect is due to the inherent nonthermal conductivity property of the impermeable medium.
When the solid porous matrix does not conduct heat, the fluid phase's ability to interact effectively with the solid phase is diminished compared to scenarios with a heat-conductive solid porous matrix.

\begin{figure}
	\centering
	\includegraphics[width=0.7\textwidth]{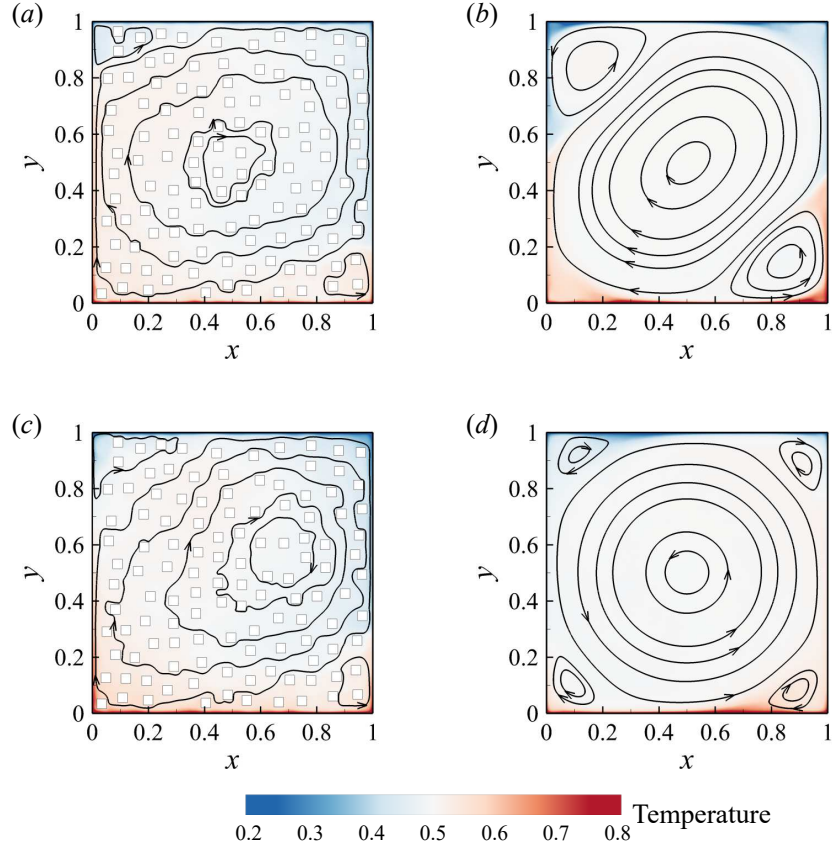}
	\caption{ Time-averaged temperature field (contour) and streamlines for (\textit{a}, \textit{b}) $Pr = 5.3$; (\textit{c}, \textit{d}) $Pr = 0.7$ in (\textit{a}, \textit{c}) a porous RB convection cell with $\phi=0.86$ and (\textit{b}, \textit{d}) a canonical RB convection cell (i.e., $\phi=1.00$).}
\label{Fig_avg_temperature}
\end{figure}

In Fig. \ref{Fig_avg_temperature}, we show the time-averaged flow field (temperature field and streamlines), where we can see that the presence of the solid porous matrix disrupts the LSC (i.e., the tilted elliptical main roll at $Pr = 5.3$ or the circular main roll at $Pr = 0.7$ in the canonical RB cell), and the LSC shape becomes more irregular in the porous RB cell.
We note the existence of solid surfaces with a no-slip boundary condition significantly affects the flow patterns, resulting in the differences in the smoothness of the streamlines between the porous RB cell and the canonical RB cell.
Meanwhile, to address concerns regarding simulation convergence, we have checked temperature fields and streamlines averaged over different time intervals to demonstrate convergence (not shown here for clarity).
The corner rolls in the porous cell are also suppressed due to the existence of the porous matrix.
Overall, we expect such disruption would lead to much more complex substructures inside the LSC at some porosities.
Specifically, when the porosity is too large, the solid porous matrix occupies only a small volume fraction in the convection cell and it will have minor effects on the flow structure; when the porosity is too small, detached plumes from the top and bottom walls will only penetrate through the porous throat, and the dense porous structure may prohibit the formation of the LSC.
Once the substructures emerge inside the LSC, they contribute to an increased instability within the LSC, potentially inducing flow reversals in turbulent thermal convection \cite{chen2019emergence}.
Our simulations have confirmed this conjecture, as we indeed observe flow reversal in the porous RB convection at $Ra = 10^{9}$ and $Pr = 0.7$ with various porosities.
Previous study by Sugiyama \emph{et al.} \cite{sugiyama2010flow} suggests that flow reversal is absent in the canonical 2D RB convection at the same $Ra$ and $Pr$ (i.e., $Ra=10^{9}$ and $Pr=0.7$),
thus the solid porous matrix has a profound impact on the flow dynamics, and understanding these effects is essential for predicting and controlling convection in porous media.

\begin{figure}
	\centering
	\includegraphics[width=0.85\textwidth]{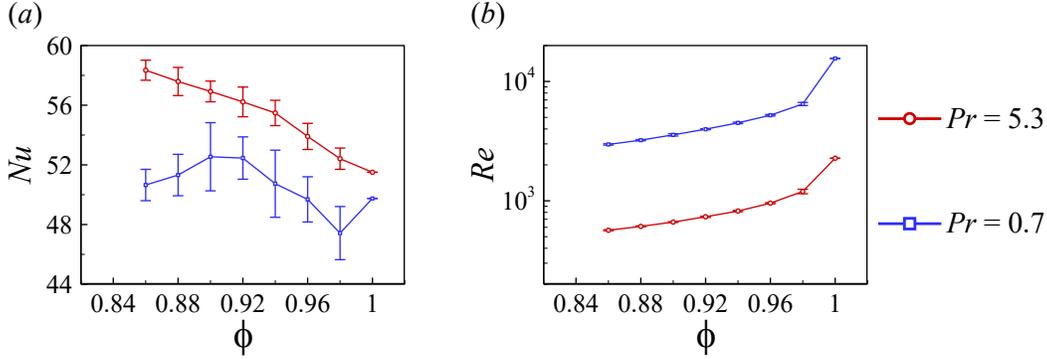}
	\caption{(\textit{a}) Nusselt number and (\textit{b}) Reynolds number as a function of porosity for $Pr = 5.3$ and $Pr = 0.7$. The error bar is calculated based on results from five different realizations of the porous structure at the same porosity.}
\label{Fig4_Nu_Re}
\end{figure}

We measure the global heat transport by the volume-averaged Nusselt number ($Nu$) as $Nu=1+\sqrt{Pr Ra}\left\langle v^{*}T^{*} \right\rangle_{V,t}$, while the global strength of the convection is measured by the Reynolds number ($Re$) as $Re=\sqrt{Ra/Pr}\sqrt{\left \langle u^{*2}+v^{*2}\right \rangle_{V,t}}$.
Here, $\langle \cdot  \rangle_{V,t}$ denotes the superficial phase and time average, the asterisk superscript (*) denote the dimensionless variables.
At each porosity, we calculate the $Nu$ and $Re$ based on results from five different realizations of the porous structure.
From Fig. \ref{Fig4_Nu_Re}(a), we can see that $Nu$ increases monotonously with the decreasing of porosity over $0.86 \le \phi \le 0.98$  at $Pr = 5.3$, while $Nu$ increases first and then decreases with the decreasing of porosity at $Pr = 0.7$.
The enhanced heat transfer efficiency with slightly decreasing porosity is attributed to strongly correlated velocity and temperature fields.
However, further decreasing the porosity increases the impedance from the porous solid matrix on heat transfer.
We hypothesize the competition of these two factors results in an optimal porosity value when heat transfer efficiency is maximized.
For $Pr = 0.7$, we observed this hypothesized optimal value at porosity around $0.90$,
while for $Pr = 5.3$, the optimal porosity value may be smaller than those under investigations (i.e., $\phi <0.86$), thus we did not observe such optimal value in our simulations.
It should be noted that to ensure adequate resolution of the pore spaces, we only consider sparse porous media in our simulations;
in addition, considering the observations alongside the presence of error bars, we recognize the necessity for caution when attempting to draw definitive conclusions regarding the existence of an optimal porosity.
This is particularly crucial when accounting for the potential influence of varying Prandtl numbers.
As for the global flow strength, we can see from Fig. \ref{Fig4_Nu_Re}(b) that $Re$ decreases monotonously with the decreasing of the porosity, which can be understood as the introduction of the porous solid matrix in the convection cell enhances flow resistance.

\begin{figure}
	\centering
	\includegraphics[width=0.8\textwidth]{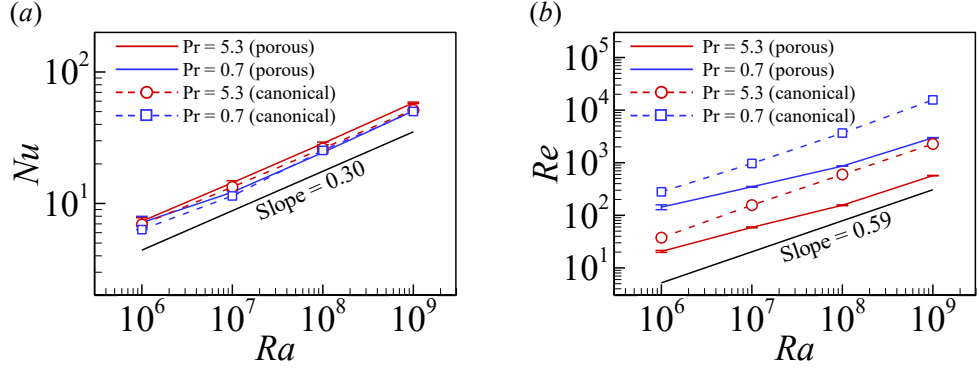}
	\caption{(\textit{a}) Nusselt number, (\textit{b}) Reynolds number as functions of Rayleigh number for two Prandtl numbers, when the porosity is fixed as $\phi = 0.86$.
The error bar is calculated based on results from five different realizations of the porous structure at the same porosity.}
\label{Figure_Nu_Ra_Phi}
\end{figure}

We also show the scaling of the global quantities, such as $Nu$ and $Re$, on one of the control parameters $Ra$ (for $10^{6} \le Ra \le 10^{9}$), while the porosity is fixed as $\phi = 0.86$.
We also provide $Nu$ and $Re$ in the canonical RB convection.
Previously, Zhang \emph{et al.} provided tabulated values of $Nu$ and $Re$ versus $Ra$ at $Pr = 5.3$ and 0.7.
Our simulation results on the canonical RB convection are in good agreement with those reported by Zhang \emph{et al.} \cite{zhang2017statistics}.
The data shown in Fig. \ref{Figure_Nu_Ra_Phi} indicate that in the porous convection, the increase of $Nu$ and $Re$ gradually approaches the power-law relations $Nu \propto Ra^{0.30}$ and $Re \propto  Ra^{0.59}$, consistent with previous results reported in the canonical RB convection \cite{van2012flow,huang2016effects,zhang2017statistics}.
At fixed $\phi$, the scaling behavior of $Nu$ and $Re$ with $Ra$ slightly deviates from that of the canonical RB convection when $Ra$ is smaller, suggesting that heat transfer and momentum exchange are not solely governed by the boundary layer.

\subsection{Statistics of temperature}

\begin{figure}
\centering
\includegraphics[width=0.7\textwidth]{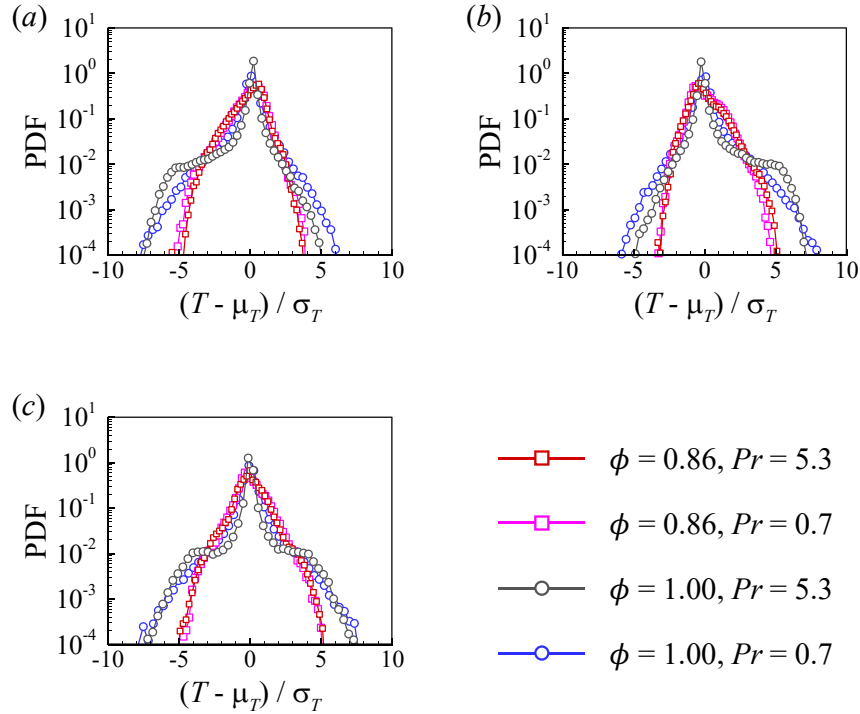}
\caption{Probability density functions (PDFs) of the normalized temperature $(T-\mu_{T})/\sigma_{T}$ measured at (\textit{a}) $y = 0.75 H$, (\textit{b}) $y = 0.25 H$, (\textit{c}) $y = 0.5 H$ in both porous and canonical RB convection cells.}
\label{Fig_PDF_T_heights}
\end{figure}

Figure \ref{Fig_PDF_T_heights} shows the probability density functions (PDFs) of normalized temperature $(T-\mu_{T})/\sigma_{T}$  measured at three different heights.
In both porous and canonical RB cells, the PDFs of temperature are left-skewed near the top region (i.e., $y = 0.75 H$) due to dominated cold falling plumes, right-skewed near the bottom region (i.e., $y = 0.25 H$) due to dominated hot rising plumes, and symmetric at mid-height (i.e., $y = 0.5 H$) as a result of comparable falling cold plumes and rising hot plumes.
Meanwhile, despite these similarities, there are notable differences between the PDF profiles of the two cells.
Specifically, in the canonical RB cell, the peak of the temperature PDF profiles exhibits a stretched exponential behavior, and the tails show a Gaussian behavior; in contrast, in the porous RB cell, the temperature PDF profiles are narrowed down, and the stretched exponential peaks are absent, indicating that porous media suppresses extreme temperature events in the convection cell.

\begin{figure}
\centering
\includegraphics[width=0.7\textwidth]{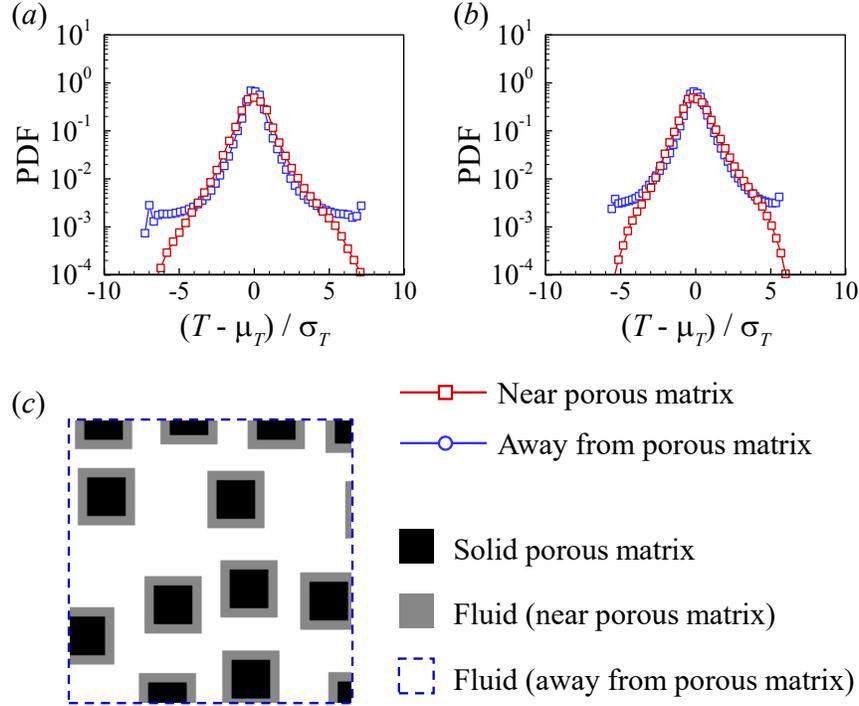}
\caption{Probability density functions (PDFs) of the normalized temperature $(T-\mu_{T})/\sigma_{T}$ measured near the porous matrix and away from the porous matrix, respectively, for (\textit{a}) $Pr=5.3$ and (\textit{b}) $Pr=0.7$, at $\phi=0.86$.
(\textit{c}) An enlarged view in the porous cells, the black region represents the porous matrix, the grey region represents the fluid near the porous matrix, and the white region represents the fluid away from the porous matrix.}
\label{Fig_PDF_T_porousRegion}
\end{figure}

To highlight the damping effect that arises from the presence of a porous structure, which impedes both hot and cold thermal plumes, we analyze the PDFs of the fluid temperature in two regions: near the solid porous matrix and away from it.
Specifically, we consider fluids nodes with distances less than 5 l.u. from the porous matrix as the near region, and fluid nodes with distances more than 5 l.u. as the away region, as illustrated in Fig. \ref{Fig_PDF_T_porousRegion}(c).
In Figs. \ref{Fig_PDF_T_porousRegion}(a) and \ref{Fig_PDF_T_porousRegion}(b), we plot the PDFs of the temperature obtained over the whole cell and over time in the above two regions.
We can see that the PDFs of temperature in both regions exhibit a symmetric peak, indicating a comparable occurrence of hot and cold plumes in those two regions.
However, the PDFs of temperature near the porous matrix have narrower tails compared to that away from the porous matrix.
This narrower tail implies a reduced degree of small-scale intermittency of the temperature fluctuations near the porous matrix.

\begin{figure}
\centering
\includegraphics[width=0.7\textwidth]{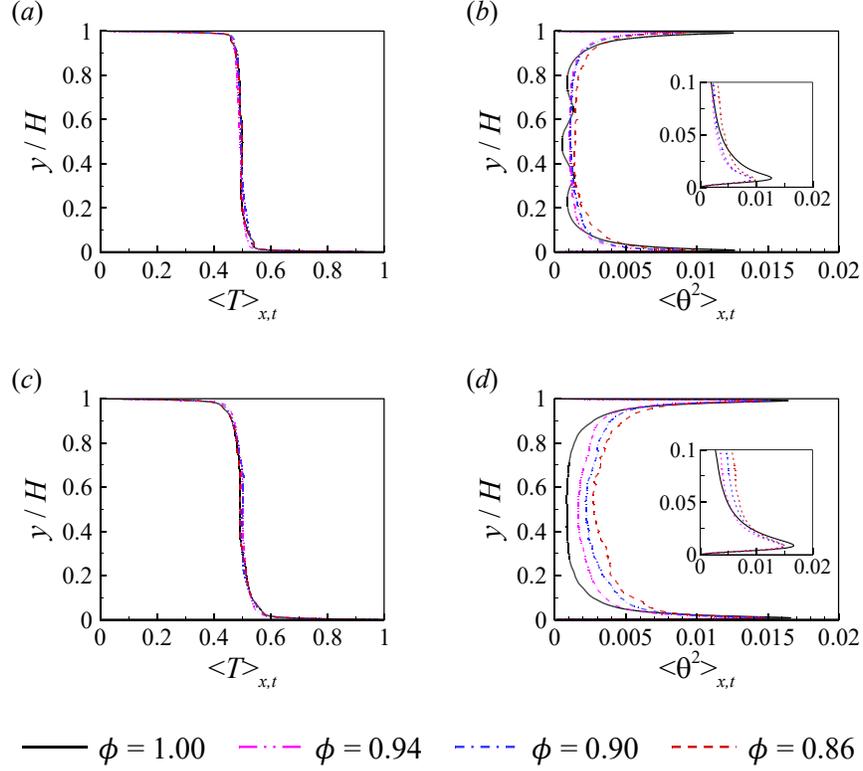}
\caption{Averaged vertical profile of (\textit{a}, \textit{c}) temperature and (\textit{b}, \textit{d}) mean-square fluctuations of temperature obtained at (\textit{a}, \textit{b}) $Pr = 5.3$ and (\textit{c}, \textit{d}) $Pr = 0.7$ for various porosities.
The vertical profiles are averaged over five different realizations of the porous structure at the same porosity.
The inset magnifies the thermal boundary layer.}
\label{Fig_horizontalT_avg_rms}
\end{figure}

We now provide the averaged vertical profile of statistics of the temperature field to quantitatively describe the temperature distributions.
We first calculate the averaged vertical profile of temperature $\langle T \rangle_{\mathbf{x},t}$ and  mean square fluctuations $\langle  \theta^{2} \rangle_{\mathbf{x},t}$  for various porosities, as shown in Fig. \ref{Fig_horizontalT_avg_rms}.
Here, the fluctuations $\theta(\mathbf{x}, t)=T(\mathbf{x},t)-\langle T \rangle_{\mathbf{x},t}(y)$; the average $\langle \cdot \rangle_{\mathbf{x},t}$  is calculated over time $t$ and along the horizontal line $\mathbf{x}$ in the fluid phase.
Note that the average over the fluid phase is referred to as the intrinsic phase average, in contrast to the superficial phase average, which would encompass the entire porous media domain.
For the porous RB convection, the vertical profiles are further averaged over five different realizations of the porous structure at the same porosity.
This fivefold averaging is conducted to mitigate the statistical errors arising from the random distribution of the porous medium.
We can see from Figs. \ref{Fig_horizontalT_avg_rms}(a) and \ref{Fig_horizontalT_avg_rms}(c), away from the top and bottom walls, the averaged vertical temperature profiles are almost a constant value of $(T_{hot}+T_{cold})/2$ in both canonical and porous RB cells.
This finding suggests that the temperature field in the bulk region of the fluid is insensitive to the presence of the impermeable solid matrix.
In contrast, Figs. \ref{Fig_horizontalT_avg_rms}(b) and \ref{Fig_horizontalT_avg_rms}(d) reveal that the averaged vertical profiles of mean-square fluctuations of the temperature are sensitive to the porosity.
With the decreasing of porosity, the flow structure becomes less coherent, leading to an increase in the fluctuations of temperature.
From the inset of Figs. \ref{Fig_horizontalT_avg_rms}(b) and \ref{Fig_horizontalT_avg_rms}(d), we can also measure the thickness of thermal boundary layer $\delta_{T}$ as the location of the peak value in the profile \cite{emran2008fine,guzman2022flow}, which is close to $H/(2Nu)$.
For the canonical RB convection (i.e., $\phi=1.00$), the temperature fluctuation profile diverges between $Pr$ of 5.3 and 0.7.
As shown in Fig. \ref{Fig_horizontalT_avg_rms}(b), the profile at $Pr = 5.3$ exhibits two peaks around $y = 0.65$ and $y = 0.35$, different from the pattern observed at $Pr = 0.7$ [see Fig. \ref{Fig_horizontalT_avg_rms}(d)].
This variation could be attributed to differences in the corner rolls at different Prandtl numbers.
Upon comparing with Fig. \ref{Fig_avg_temperature}(b), it becomes apparent that the enhancement in turbulent fluctuations coincides with heights of corner rolls that are situated diagonally, thereby leading to the emergence of the peaks observed in the bulk region of temperature fluctuation profile at $Pr = 5.3$.
These differences highlight the complexity of the system's temperature fluctuations and plume dynamics in response to the Prandtl number.

\begin{figure}
\centering
\includegraphics[width=0.7\textwidth]{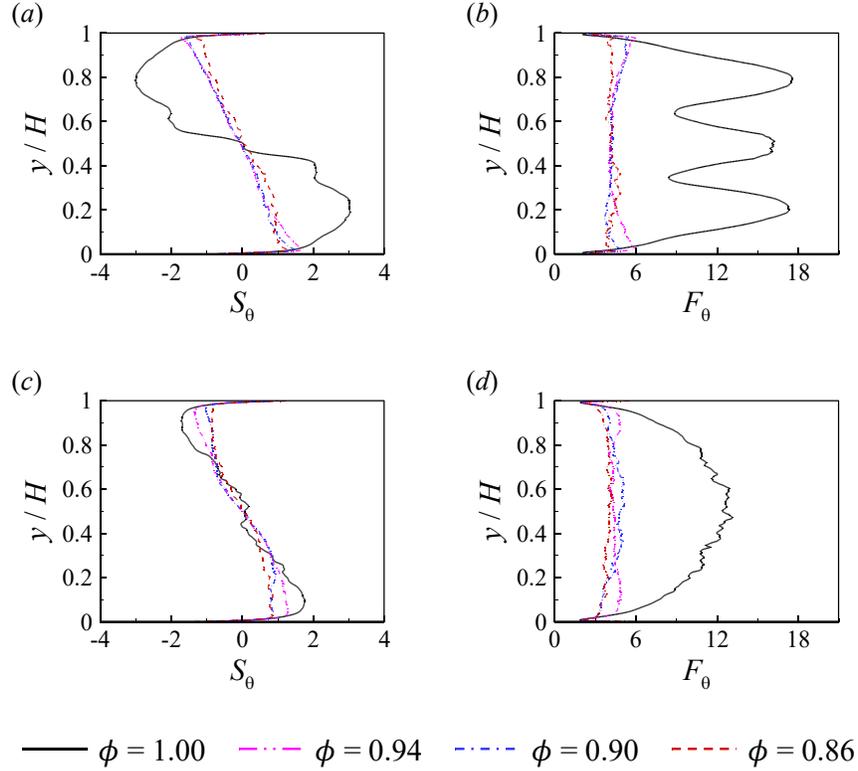}
\caption{Averaged vertical profile of (\textit{a}, \textit{c}) skewness of temperature and (\textit{b}, \textit{d}) flatness of temperature obtained at (\textit{a}, \textit{b}) $Pr = 5.3$ and (\textit{c}, \textit{d}) $Pr = 0.7$ for various porosities.}
\label{Fig_horizontalT_ske_fla}
\end{figure}

We further calculate the averaged vertical profile of higher-order moments of temperature and plot the skewness $S_{\theta}(y)=\langle \theta^{3} \rangle_{\mathbf{x},t}/\langle \theta^{2} \rangle_{\mathbf{x},t}^{3/2}$ and flatness $F_{\theta}(y)=\langle \theta^{4} \rangle_{\mathbf{x},t}/\langle \theta^{2} \rangle_{\mathbf{x},t}^{2}$ of temperature, as shown in Fig. \ref{Fig_horizontalT_ske_fla},
where the vertical profiles are averaged over five different realizations of the porous structure at the same porosity.
Skewness evaluates the asymmetry of the distribution, whereas flatness quantifies the extent of the distribution's tails and reveal how extreme values deviate from the mean.
Compared to the porous RB cell, the skewness has smaller absolute values near the top (bottom) regions in the canonical RB cell, indicating that the localized cold falling (hot rising) plumes has more profound effects in the canonical RB cell.
On the other hand, from Figs. \ref{Fig_PDF_T_heights}(a) and \ref{Fig_PDF_T_heights}(b), we can also observe that the temperature PDF is more symmetric for the porous convection at the heights of $0.25H$ and $0.75H$, which aligns with the lower skewness values seen for temperature in porous convection.
In both canonical and porous RB cells, the skewness values are around zero at midheight of $y = 0.5 H$ [see Figs. \ref{Fig_horizontalT_ske_fla}(a) and \ref{Fig_horizontalT_ske_fla}(c)], indicating almost equal number of hot and cold plumes flow through the midheight.
We can also find that the flatness has much smaller values in the porous cell than that in the canonical RB cell, as shown in Figs. \ref{Fig_horizontalT_ske_fla}(b) and \ref{Fig_horizontalT_ske_fla}(d).
Specifically, there is a shift towards a Gaussian distribution in the temperature PDF of porous convection as compared to canonical RB convection, because solid porous matrix impedes both hot and cold thermal plumes, there are fewer fluids with temperature that deviate from the bulk temperature.
In the canonical RB convection, the differences in skewness and flatness at two Prandtl numbers can be attributed to the shape of large-scale circulation.
The LSC is in the form of a tilted ellipse at $Pr = 5.3$, occupying a diagonal position within the convection cell, with two secondary corner rolls sited along the opposing diagonal.
On the other hand, at $Pr = 0.7$, the LSC appears to be circular, with four secondary corner vortices present.
This elliptical arrangement of the LSC at $Pr = 5.3$ results in asymmetric hot (or cold) plumes falling back to the bottom (or top) plate, which is associated with countergradient heat transfer, thereby creating a more asymmetric temperature PDF, increasing the skewness.
At the heights of countergradient heat transfer, there is an abundance of cold and hot plumes with temperatures deviating from the bulk temperature, which are found along the edges of the corner rolls, leading to the peaks in the temperature flatness profile observed at heights of $0.2H$ and $0.8H$.

\begin{figure}
\centering
\includegraphics[width=0.65\textwidth]{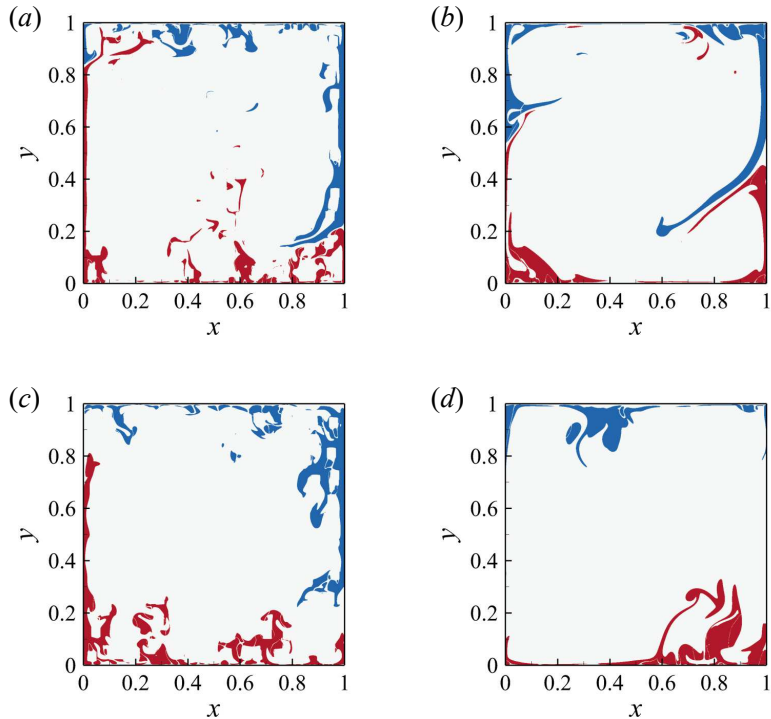}
\caption{A typical snapshot of the instantaneous plume field for (\textit{a}, \textit{b}) $Pr = 5.3$; (\textit{c}, \textit{d}) $Pr = 0.7$ in (\textit{a}, \textit{c}) a porous RB convection cell with $\phi = 0.86$ and (\textit{b}, \textit{d}) a canonical RB convection cell (i.e., $\phi = 1.00$). Here, the blue areas corresponding to cold plumes and the red areas corresponding to hot plumes.}
\label{Fig_instant_plume}
\end{figure}

To evaluate the spatial and temporal distributions of thermal plumes, we adopt the criteria similar to those used in Ref. \cite{huang2013confinement,van2015plume,zhang2017statistics}, specifically
\begin{equation}
|T(x,y,t)-\langle T \rangle_{x,t}|> cT_{rms}, \ \ \ \sqrt{Pr Ra}|v(x,y,t)T(x,y,t)|>cNu
\end{equation}
Here, $c$ is an empirical constant, for which a value of $c = 1$ is chosen.
This criterion assumes that plumes occur in regions of local temperature extremes (either maximum or minimum), and in areas where local convective heat flux is larger than the spatial and temporal averaged one.
The applicability of this empirical criterion in accurately extracting plume structures within both canonical and porous convection is evident from Fig. \ref{Fig_instant_plume}.

\begin{figure}
\centering
\includegraphics[width=0.85\textwidth]{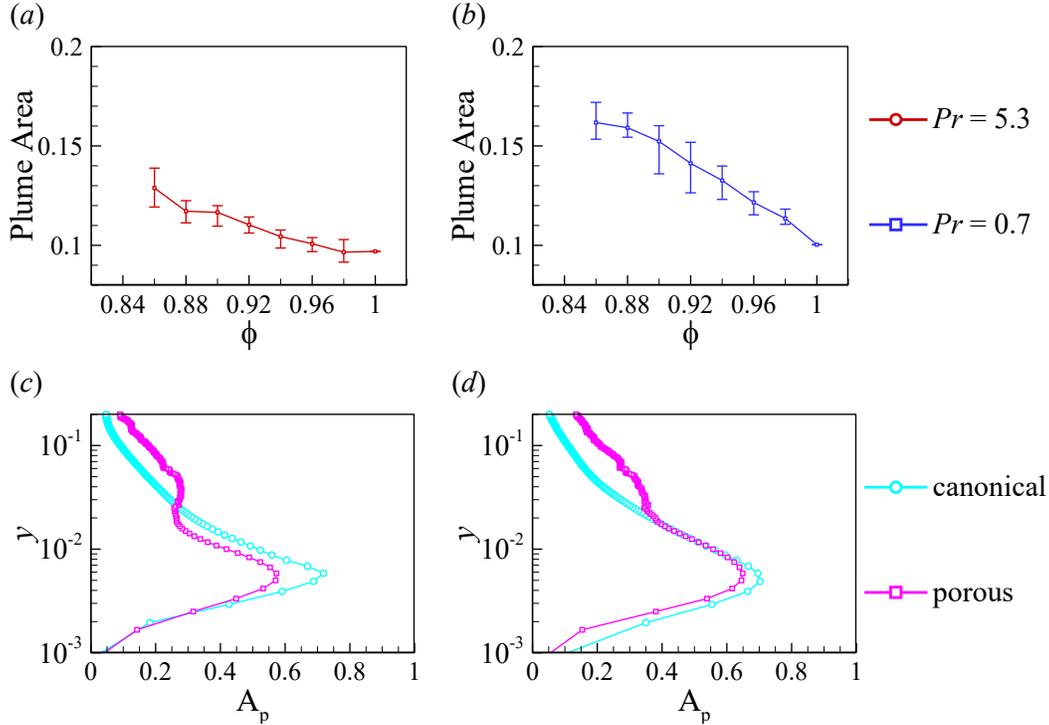}
\caption{Time-averaged plume area in the cell as functions of porosity for (\textit{a}) $Pr = 5.3$ and (\textit{b}) $Pr  = 0.7$.
The error bar is calculated based on results from five different realizations of the porous structure at the same porosity.
The averaged vertical profile of plume areas for ($\textit{c}$) $Pr = 5.3$ and ($\textit{d}$) $Pr = 0.7$.}
\label{Figure_plume-height}
\end{figure}

We calculate the time-averaged plume area within the cell, and plot the plume areas as functions of porosity.
From Figs. \ref{Figure_plume-height} (a) and \ref{Figure_plume-height}(b), we can see that with the decrease of porosity, plume areas generally increase under both Prandtl number conditions.
Additionally, we calculate the plume area along a horizontal line in the fluid phase, and we plot the averaged vertical profile near the bottom wall in Figs. \ref{Figure_plume-height}(c) and \ref{Figure_plume-height}(d).
These profiles were averaged over five different realizations of the porous structure with the same porosity.
Just above the thermal boundary layers ($y \gtrapprox 0.01H$), we observed more hot plumes within the porous convection cell compared to the canonical cell at the same height.
This observation serves as another evidence indicating that more hot plumes penetrate the pore throat after detaching from the thermal boundary layers.

\subsection{Statistics of thermal energy dissipation rate}

\begin{figure}
\centering
\includegraphics[width=0.65\textwidth]{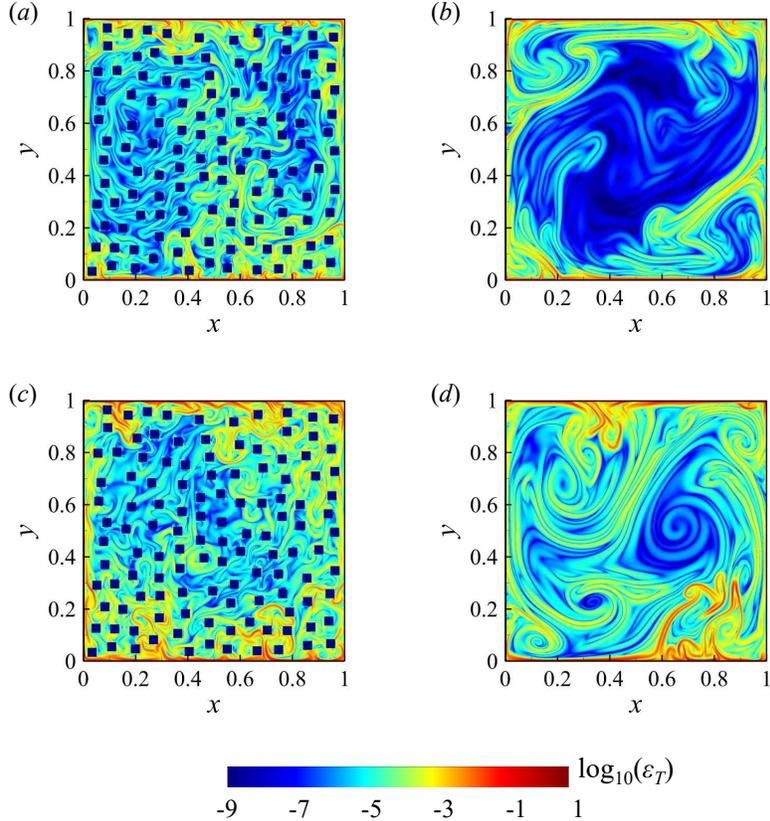}
\caption{A typical snapshot of the instantaneous logarithmic thermal energy dissipation rate field for (\textit{a}, \textit{b}) $Pr = 5.3$; (\textit{c}, \textit{d}) $Pr = 0.7$ in (\textit{a}, \textit{c}) a porous RB convection cell with $\phi=0.86$  and (\textit{b}, \textit{d}) a canonical RB convection cell (i.e., $\phi=1.00$).}
\label{Fig_instant_EpsilonT}
\end{figure}

A typical snapshot of an instantaneous logarithmic thermal energy dissipation rate field in both the porous and the canonical RB cell is shown in Fig. \ref{Fig_instant_EpsilonT}.
Overall, we can observe intense thermal energy dissipations occur near the top and bottom boundary layers, where falling cold plumes or rising hot plumes detach from the boundary layers.
Besides, we can also observe intense thermal energy dissipation in the bulk region of the porous RB cell, which is absent in the canonical RB cell.
The reason behind this observation lies in the permeability of the porous media.
In the porous RB cell, plumes can penetrate through the pore throat, leading to the formation of thermal plumes associated with large amplitudes of thermal energy dissipation rates.
It should be noted that we assume the solid porous matrix is impermeable to heat flux, thus the thermal energy dissipation rate $\varepsilon_{T}(\mathbf{x})$  is zero at the location of the solid porous matrix, leading to different distributions of  $\varepsilon_{T}(\mathbf{x})$  compared to previous study \cite{liu2020rayleigh}.

\begin{figure}
\centering
\includegraphics[width=0.65\textwidth]{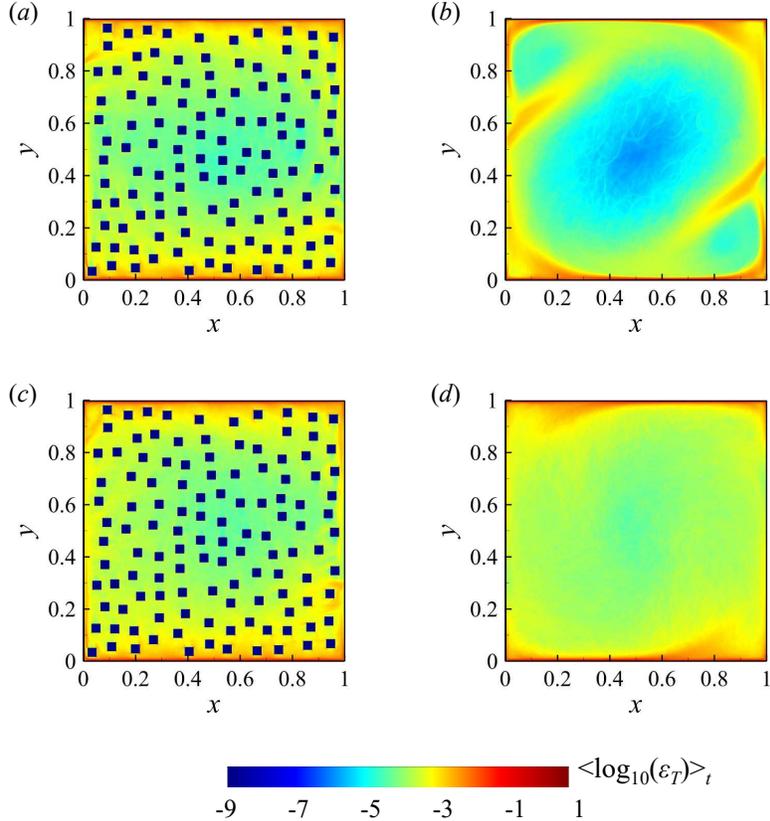}
\caption{Time-averaged logarithmic thermal energy dissipation rate field for (\textit{a}, \textit{b}) $Pr = 5.3$; (\textit{c}, \textit{d}) $Pr = 0.7$ in (\textit{a}, \textit{c}) a porous RB convection cell with $\phi=0.86$  and (\textit{b}, \textit{d}) a canonical RB convection cell (i.e., $\phi=1.00$).}
\label{Fig_avg_EpsilonT}
\end{figure}

We further show the time-averaged logarithmic thermal energy dissipation rate field in Fig. \ref{Fig_avg_EpsilonT}.
We can see that the contribution of thermal plumes to thermal energy dissipation is filtered out in both canonical and porous RB convection cells.
In the time-averaged field, we only observe intense thermal energy dissipation occurs near the top and bottom walls, as well as the edge of LSC, where strong temperature gradients exist.
Particularly, in the porous cell, we did not observe a significant increase in the thermal energy dissipation rate at the fluid-solid interface of the porous matrix, in contrast to the previous study \cite{liu2020rayleigh}.

\begin{figure}
\centering
\includegraphics[width=0.65\textwidth]{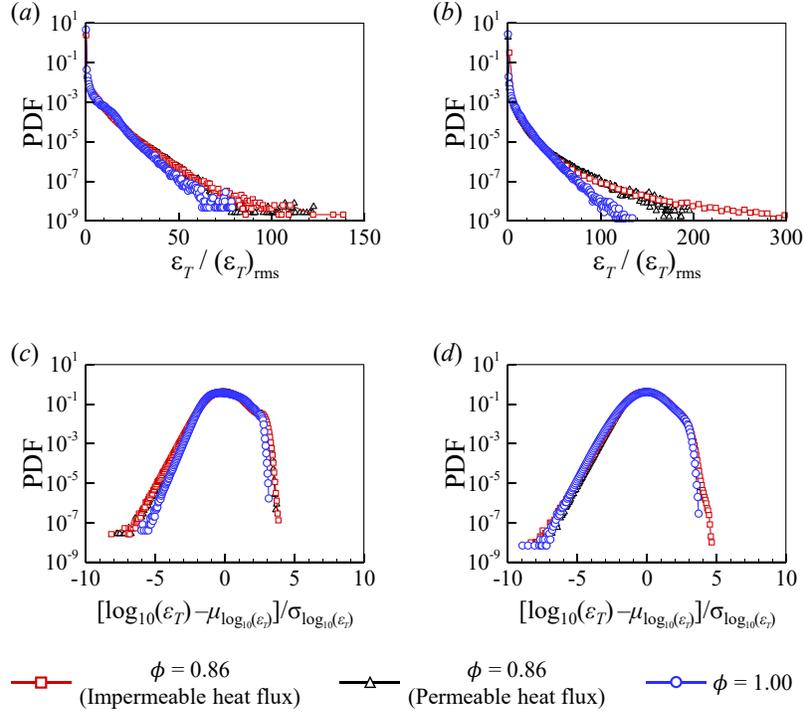}
\caption{ (\textit{a}, \textit{b}) Probability density functions (PDFs) of the thermal energy dissipation rate $\varepsilon_{T}(\mathbf{x},t)$, and (\textit{c}, \textit{d}) PDFs of the normalized logarithmic thermal energy dissipation rate $\log_{10}\varepsilon_{T}(\mathbf{x},t)$  obtained over the whole fluid region in the cell at (\textit{a}, \textit{c}) $Pr = 5.3$ and (\textit{b}, \textit{d}) $Pr = 0.7$.}
\label{Fig_PDF_epsilonT}
\end{figure}

We plot the PDFs of thermal energy dissipation rates $\varepsilon_{T}(\mathbf{x}_f,t)$  obtained over the fluid phase in the cell over time, further normalized by their root-mean-square (rms) values, as shown in Figs. \ref{Fig_PDF_epsilonT}(a) and \ref{Fig_PDF_epsilonT}(b).
Compared with the canonical RB convection, the PDF tails of porous RB convection are more extended, indicating an increasing degree of small-scale intermittency in the thermal energy dissipation field due to the presence of the porous solid matrix.
In the figures, we also compared the PDFs of thermal energy dissipation rates with the solid porous matrix being either permeable or impermeable to heat flux.
We observed that when the porous matrix is impermeable to heat flux (represented by the red squares), the tails of the PDFs are longer.
However, in both scenarios, the tails of the PDFs for thermal energy dissipation rate exceed those found in canonical RB convection.
This observation suggests that while the solid porous matrix generally enhance small-scale intermittency, the thermal physical property of the porous matrix also influence the behavior of small-scale intermittency.
The PDF of the scalar dissipation rate plays a crucial role in describing turbulent isothermal and reacting flows.
It is common to use a log-normal PDF to characterize the distribution of dissipation rate values \cite{bilger2004some}.
We further check whether the thermal energy dissipation fields in the porous RB convection follow a log-normal distribution or a non-log-normal distribution as observed in previous studies of canonical RB convection \cite{zhang2017statistics,xu2019statistics}.
In Figs. \ref{Fig_PDF_epsilonT}(c) and \ref{Fig_PDF_epsilonT}(d), we plot the PDFs of the normalized logarithmic thermal energy dissipation rate $\left(\log_{10} \varepsilon_{T}-\mu_{\log_{10} \varepsilon_{T}}\right) / \sigma_{\log_{10} \varepsilon_{T}}$.
We can observe clear departures from log normality of the thermal energy dissipation field for both canonical and porous RB convections, as a result of intermittent local dissipation.
Thus, we conjecture that the non-log-normal distribution for thermal energy dissipation rate is universal for buoyancy-driven turbulent convection, even in the presence of complex flow geometry.

\section{\label{Section4}Conclusions}

In this work, we have conducted pore-scale direct numerical simulations of thermal convective flow at vigorous convection regime (i.e., the porous Rayleigh number range $10^{5} < Ra^{*} < 10^{7}$) \cite{hewitt2020vigorous}.
Our simulation results showed that the solid porous matrix, which is impermeable to both fluid and heat flux, significantly impacts the plume dynamics in the porous RB cell.
In the porous RB convection, compared to the case of solid porous matrix being permeable to heat flux, the plume dynamics are less coherent when the solid porous matrix is impermeable to heat flux.

Furthermore, we investigated the statistical properties of temperature and thermal energy dissipation rate in the porous RB cell.
We found that the averaged vertical temperature profiles are almost a constant value, regardless of the porosity of the cell.
However, as the porosity decreases, the mean-square fluctuations of temperature increases, and the absolute values of skewness and flatness are much smaller in the porous RB cell compared to the canonical RB cell.
This indicates that the flow is less turbulent in the porous media.

Our study also revealed that intense thermal energy dissipation occurs near the top and bottom walls, as well as in the bulk region of a porous RB cell.
We observed that the small-scale thermal energy dissipation field is more intermittent in the porous cell compared to the canonical RB cell.
Despite this difference, both cells exhibit a non-log-normal distribution of thermal energy dissipation rate.

In summary, our pore-scale direct numerical simulations of porous thermal convective flow provide important insights into the behavior of coupled fluid flow and heat transfer in porous media.
Our findings highlight the impact of the solid porous matrix on the plume dynamics, temperature profiles, and thermal energy dissipation rate, which are crucial for the development of more accurate REV-scale models \cite{gasow2021macroscopic}.

\begin{acknowledgments}
This work was supported by the National Natural Science Foundation of China (NSFC) through Grants No. 12272311 and No. 12125204,
and the 111 project of China (Project No. B17037).
The authors acknowledge the Beijing Beilong Super Cloud Computing Co., Ltd for providing HPC resources that have contributed to the research results reported within this paper.
\end{acknowledgments}

\appendix

\section{Simulation details of porous convection}
We provide simulation results at Prandtl numbers of $Pr = 5.3$ and $0.7$ and a fixed Rayleigh number of $Ra = 10^{9}$, the porosity $\phi$ range  $0.86 \leq \phi \le 0.98$.
In addition, we vary the $Ra$ for $10^{6} \le Ra \le 10^{9}$ at two fixed Pr, while $\phi$ is fixed as 0.86.
Thus, a total of 100 simulations were carried for porous convection with impermeable solid matrix, and tabulated values on the results are listed in Table \ref{tab:addlabel}.

\newpage
\setlength{\floatsep}{1pt} 
\setlength{\textfloatsep}{1pt} 
\begin{table}
	\begin{center}
		\begin{minipage}{0.8\linewidth}
        \caption{Simulation details of porous convection.
				The columns from left to right indicate the following: the Rayleigh number $Ra$, the Prandtl number $Pr$, the grid numbers, the cylinder length $d$, the cylinder number $N_{d}$, the porosity $\phi$, the Nusselt number $Nu$ for five different realizations of porous structure as well as its mean value and standard deviation.}\label{tab:addlabel}
		\end{minipage}
	\end{center}
\end{table}
\begin{longtable}[htbp]{ccccccccccccccccccccccccc}
	\toprule
	& & & & & & \multicolumn{5}{c}{Nusselt Number ($Nu$)} \\
	\cmidrule{7-11}
	$Ra$ & $Pr$ & Grids & $d$ & $N_d$ & $\phi$ & Case\#1 & Case\#2 & Case\#3 & Case\#4 & Case\#5 \\
	\midrule
	\endfirsthead

	$10^9$ & 5.3 & {1200×1200} & 40 & 126 & 0.86 & 58.69 & 57.07 & 58.09 & 58.85 & 58.80 \\
	          &       &       &       & {108} & {0.88} & 57.50  & 58.28 & 58.20  & 58.22 & 56.58 \\
	          &       &       &       & {90} & {0.9} & 56.18 & 57.18 & 57.59 & 58.05 & 56.13 \\
	          &       &       &       & {72} & {0.92} & 54.95 & 56.30  & 55.64 & 56.84 & 57.45 \\
	          &       &       &       & {54} & {0.94} & 54.84 & 55.45 & 55.54 & 54.38 & 56.70 \\
	          &       &       &       & {36} & {0.96} & 54.17 & 54.62 & 54.01 & 52.59 & 53.79 \\
	          &       &       &       & {18} & {0.98} & 52.12 & 52.18 & 53.91 & 51.76 & 52.42 \\
	{$10^9$} & {0.7} & {{1200×1200}} & {40} & {126} & {0.86} & 52.13 & 50.48 & 51.11 & 49.29 & 50.20 \\
	          &       &       &       & {108} & {0.88} & 49.83 & 52.15 & 53.29 & 50.78 & 50.77 \\
	          &       &       &       & {90} & {0.9} & 50.22 & 53.82 & 56.19 & 52.47 & 51.43 \\
	          &       &       &       & {72} & {0.92} & 51.95 & 54.79 & 52.62 & 51.21 & 51.48 \\
	          &       &       &       & {54} & {0.94} & 49.29 & 54.45 & 51.27 & 50.92 & 48.14 \\
	          &       &       &       & {36} & {0.96} & 50.44 & 51.70  & 49.29 & 49.07 & 47.71 \\
	          &       &       &       & {18} & {0.98} & 44.65 & 46.80  & 48.17 & 48.74 & 49.08 \\
	{$10^8$} & {5.3} & {{600×600}} & {20} & {126} & {0.86} & 28.66 & 28.61 & 29.02 & 29.29 & 27.57 \\
	{$10^8$} & {0.7} & {{600×600}} & {20} & {126} & {0.86} & 24.58 & 24.14 & 24.44 & 24.83 & 24.13 \\
	{$10^7$} & {5.3} & {{300×300}} & {20} & {32} & {0.86} & 14.04 & 14.56 & 14.25 & 13.94 & 15.18 \\
	{$10^7$} & {0.7} & {{300×300}} & {20} & {32} & {0.86} & 11.85 & 12.11 & 12.07 & 12.28 & 12.64 \\
	{$10^6$} & {5.3} & {{150×150}} & {20} & {8} & {0.86} & 7.18  & 7.55  & 7.98  & 6.74  & 7.20 \\
	{$10^6$} & {0.7} & {{150×150}} & {20} & {8} & {0.86} & 7.02  & 7.71  & 6.72  & 6.12  & 8.23 \\
				\bottomrule
\end{longtable}


%

\end{document}